\documentclass[10pt]{iopart}
\usepackage{graphicx}
\usepackage{cite}
\usepackage{array}
\newcolumntype{?}{!{\vrule width 1pt}}

\begin{document}
	
\title{Auger recombination of dark excitons in  ${\bf WS_2}$ and ${\bf WSe_2}$ monolayers}

\author{Mark Danovich, Viktor Z\'{o}lyomi, Vladimir I. Fal'ko}
\address{National Graphene Institute, University of Manchester, Booth St E, Manchester M13 9PL, UK}
\author{Igor L. Aleiner}
\address{Physics Department, Columbia University, New York, NY 10027, USA}



\begin{abstract}
	We propose a novel phonon assisted Auger process unique to the electronic band structure of monolayer transition metal dichalcogenides (TMDCs), which dominates the radiative recombination of ground state excitons in Tungsten based TMDCs. Using experimental and DFT computed values for the exciton energies, spin-orbit splittings, optical matrix element, and the Auger matrix elements, we find that the Auger process begins to dominate at carrier densities as low as $10^{9\--10}~{\rm cm^{-2}}$, thus providing a plausible explanation for the low quantum efficiencies reported for these materials.
\end{abstract}

\ead{mark.danovich@postgrad.manchester.ac.uk}

\maketitle
\ioptwocol

Recently, there was an expansive interest in Monolayer transition metal dichalcogenides (TMDCs) due to their potential in optoelectronic applications \cite{devices, tmdcs_opto, cao}. In contrast to bulk TMDC crystals, the monolayers of ${\rm MoS_2, MoSe_2, WS_2},$ and ${\rm WSe_2}$ are direct band semiconductors. Contrary to ${\rm III-V}$ semiconductors, in these hexagonal 2D crystals the conduction ($c$) and valence ($v$) bands edges are at the $K/K'$ points of the Brillouin zone (BZ) rather than at the $\Gamma$-point. Several experiments have alre-ady demonstrated a strong light-matter interaction in these 2D crystals\cite{strong}.

Potentially practical implementations of these TMDC atomic crystals in optoelectronic devices require high quantum efficiency of the optical process. However, despite the recent progress in improving the quality of 2D TMDCs, the quantum efficiency observed in photoluminescence experiments\cite{mak,plmos2,wang_auger} never exceeded 1\%. 
Such systematically low quantum efficiency calls for finding the mechanism responsible for the non-radiative recombination of electron-hole pairs, excitons, or trions.

Here we show that there exists a phonon assisted Auger recombination process illustrated in Fig.~1, which is specific for 2D TMDC semiconductors. By explicit comparison of the phonon assisted radiative recombination rate of ground state excitons in monolayers of ${\rm WS_2}$ and ${\rm WSe_2}$ with the rate due to the suggested Auger mechanism, we find that the latter starts dominating at electron densities as low as $10^{9\--10}~{\rm cm}^{-2}$. 
The specific band structure of 2D TMDCs defies the common wisdom (based on III-V semiconductors studies) that the 2D confinement quenches Auger recombination processes. Namely, electrons from the vicinity of the conduction band ($c$) edge, can undergo a transition into one of the higher bands ($c'$ in Fig.~1) which is almost in resonance with the exciton annihilation. Both $c$ and $c'$ bands are comprised of $d$-orbitals of the same metal atom, facilitating the Auger transition.

\begin{figure}[!t]
	\centering
	\includegraphics[width=0.46\textwidth]{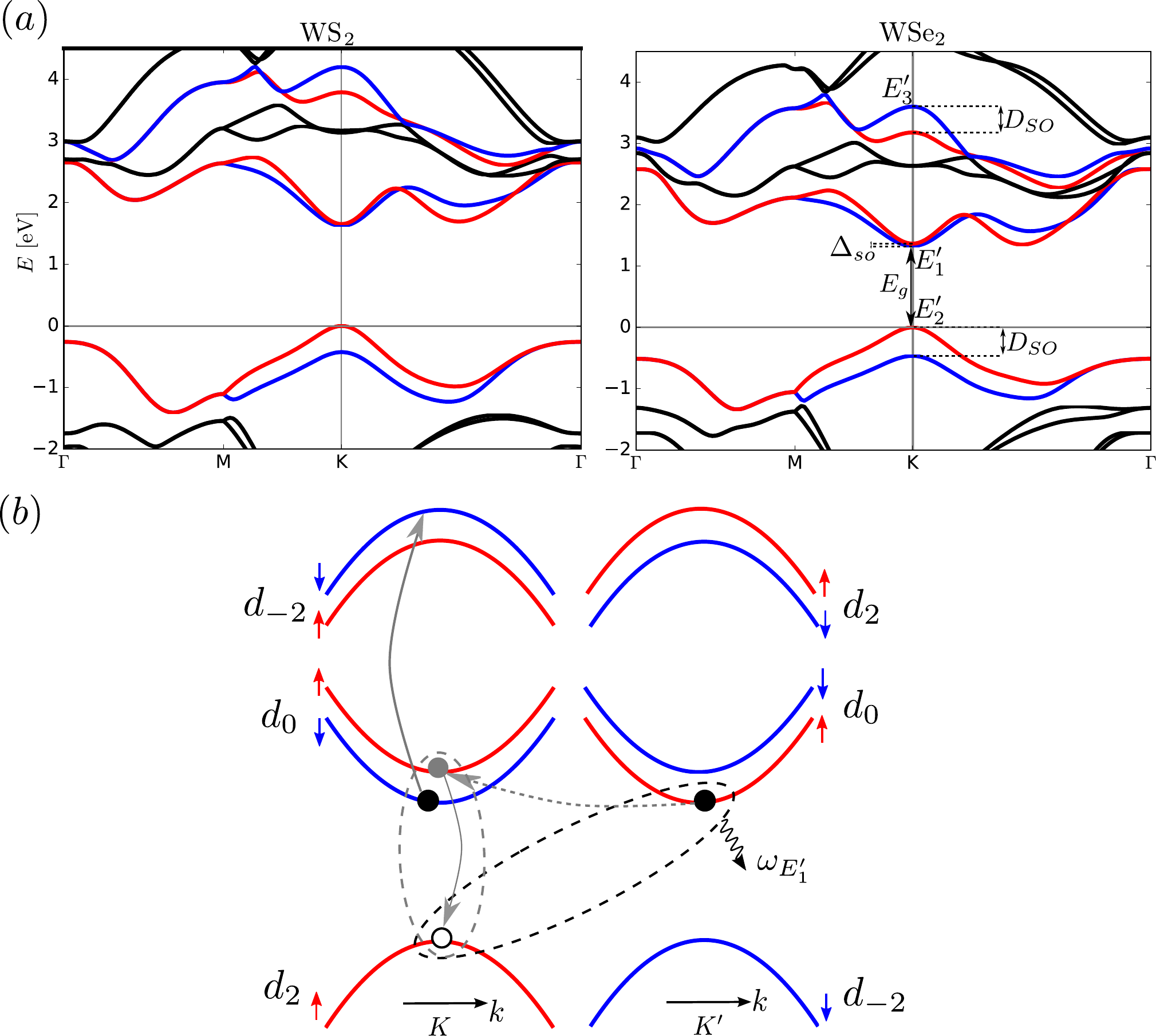}
	\caption{(a) DFT calculated \cite{QE-2009} band structure of ${\rm WS_2}$ and ${\rm WSe_2}$.
	(b) Sketch of the electronic band structure near the $K/K'$ points, and schematics of the phonon assisted Auger process. The dashed gray line corresponds to the virtual transition.} 
	\label{fig:fig_intro}
\end{figure}

The electron band structure near the corners of the BZ is described by
\begin{equation}
\hat{H}_0=\sum_{\nu\sigma\tau}\int d^2\vec{r} \Psi^{\dagger}_{\nu\sigma\tau}\epsilon_{\nu\sigma\tau}(-i\hbar \nabla)\Psi_{\nu\sigma\tau},
\end{equation}
where $\nu=v,c,c', \quad \sigma=\pm(\uparrow,\downarrow),$ and $\tau=\pm(K,K')$.
With reference to the band structure shown in Fig.~1, the relevant spectrum of electrons in the vicinity of the BZ corners can be described using the effective mass approximation as $\epsilon_{\nu}=E_{\nu\sigma\tau}+\frac{\hbar^2k^2}{2m_{\nu}}$.
Here we count energies from the $v$ band edge and 
$E_{v}=-\frac{D_{SO}}{2}(1-\tau\sigma)$, with $m_{v}<0$. For the conduction bands we use 
$E_{c}=E_g+\frac{\Delta_{SO}}{2}(1+\tau\sigma)$, and $E_{c'}=2E_g+\Upsilon-\frac{D_{SO}}{2}(1+\tau\sigma)$.
Due to mirror plane symmetry ($\sigma_h$) of 2D TMDCs, the electron spin projection on to the z-axis normal to the plane $\sigma$, is a good quantum number. The signs of spin-orbit splittings in $E_c$ reflect the inverted order of spin-split states in c and v bands specific to Tungsten based TMDCs\cite{kdotp, wang} (in contrast to their Molybdenum counterparts). This results in a ground state exciton/trion which is dark due to spin and momentum conservation constraints\cite{heinz}, requiring emission of $K$-point phonons for recombination.

\begin{table*}[!t]
	\centering
	\caption{Character table for the irreducible representations of the extended point group $C_{3v}''$, and their correspondence to the relevant fermionic and bosonic fields.}
	\label{c3v''}
	\begin{tabular}{lcccccc?c}
		\br
		$C_{3v}''$ & $E$ & $t,t^2$ & $2C_3$ & $9\sigma_v$ & $2tC_3$ & $2t^2C_3$ & \\
		\mr
		$A_1$ & 1 & 1 & 1 & 1 & 1 & 1 & \\ 
		$A_2$ & 1 & 1 & 1 & -1 & 1 & 1  & \\ 
		$E$ & 2 & 2 & -1 & 0 & -1 & -1 & $(\mathcal{E}_x,\mathcal{E}_y)$ \\ 
		\hline
		$E_1'$ & 2 & -1 & -1 & 0 & 2 & -1 & $\Psi_c$ \\ 
		$E_2'$ & 2 & -1 & 2 & 0 & -1 & -1 & $\Psi_v$  \\ 
		$E_3'$ & 2 & -1 & -1 & 0 & -1 & 2 & $\Psi_{c'}$	\\
		\br
		$D_{xy}$ & 12 & 0  & 0  & 0 & -3 & -3  & ${\rm phonons}$  \\
		$D_z$ & 3  & 0  & 0  & 1 & 3 & 0 &  $b$ \\
		\br
	\end{tabular} 
\end{table*}

\begin{table*}[!t]
	\centering
	\caption{Product table for the irreducible representations of the extended point group $C_{3v}''$.}
	\label{prod}
	\begin{tabular}{lcccccc}
	\br
		$C_{3v}''$ & $A_1$ & $A_2$ & $E$ & $E_1'$ & $E_2'$ & $E_3'$\\
		\mr
		$A_1$ & $A_1$ & $A_2$ & $E$ & $E_1'$ & $E_2'$ & $E_3'$ \\ 
		$A_2$ &  $A_2$& $A_1$ & $E$ & $E_1'$ & $E_2'$ & $E_3'$ \\ 
		$E$ & $E$ & $E$ & $A_1\oplus A_2\oplus E$ & $E_2'\oplus E_3'$ & $E_1'\oplus E_3'$ & $E_1'\oplus E_2'$ \\ 
		\mr
		$E_1'$ &$E_1'$ & $E_1'$ & $E_2'\oplus E_3'$& $A_1\oplus A_2\oplus E_1'$ & $E\oplus E_3'$ & $E\oplus E_2'$ \\ 
		$E_2'$ &$E_2'$& $E_2'$ & $E_1'\oplus E_3'$ & $E\oplus E_3'$ & $A_1\oplus A_2\oplus E_2'$ & $E\oplus E_1'$ \\ 
		$E_3'$ & $E_3'$ & $E_3'$ & $E_1'\oplus E_2'$ & $E\oplus E_2'$ & $E\oplus E_1'$ & $A_1\oplus A_2\oplus E_3'$\\
		\br
	\end{tabular} 
\end{table*}
To classify suitable options for the radiative and non-radiative transitions in 2D ${\rm WS_2}$ and ${\rm WSe_2}$, we analyze its symmetry group and write down the corresponding terms in the Hamiltonian.
	The point group of 2D TMDCs is $D_{3h}$, which is a direct product group, $C_{3v}\otimes \sigma_{h}$, where $\sigma_{h}$ is the horizontal mirror reflection. The states belonging to the $v$, $c$, and $c'$ bands near the $K/K'$-points are composed of the $d_0, d_2$ and $d_{-2}$ metal orbitals which posses $z\rightarrow-z$ symmetry\cite{wang}, and therefore belong to the identity irreducible representation (irrep) of $\sigma_{h}$. As a result, we can focus on the point group $C_{3v}$ for the classification of the electronic states into irreps, as well as the classification of phonon modes coupling to the electrons states. Since the states at the $K$ and $K'$-points are degenerate, it is advantageous to treat them simultaneously. This is achieved by tripling the unit cell, resulting in a three times smaller Brillouin zone in which the $K$ and $K'$ points are folded into the $\Gamma$-point\cite{basko}. Tripling of the unit cell is achieved by factoring out two translations from the space group of the crystal resulting in the new point group $C_{3v}''=C_{3v}+tC_{3v}+t^2C_{3v}$, where $t$ denotes translation by a lattice vector, and $t^3=1$. The character table of the new point group containing 18 elements and 6 irreps is given in Table~\ref{c3v''}. In the same table we list the electron and photon fields corresponding to the irreps. The decomposition of the direct products of irreps is shown in Table~\ref{prod}. 

Using Table~\ref{c3v''} one can write down the Hamiltonian for the interaction of the electrons with light\cite{photo_gaas,excitons},
\begin{eqnarray}
\label{eq:rad} 
\fl H_{r}=\frac{e\hbar v}{E_g}\sum_{\sigma,\tau}\int d^2 \vec{r}\, \Psi^{\dagger}_{c\sigma\tau}\Psi_{v\sigma\tau}(\mathcal{E}_x+i\tau\mathcal{E}_y)+h.c.,
\end{eqnarray}
where $e$ is the electron charge, $v$ is the velocity originating from the off-diagonal momentum matrix element, and $\vec{\mathcal{E}}$ is the electric field of light.
\begin{table*}[!t]
	\caption{Material parameters used for the rates calculation. }
	\label{data}
			\centering
	\begin{tabular}{lccccccccc}
		\br
		& $\frac{m_c}{m}^{\rm a}$ &  $\frac{m_v}{m}^{\rm a}$& $\frac{m_{c'}}{m}$ & $\Delta_{SO}$ & $D_{SO}$ & $E_g^{\rm b}$ & $\Upsilon$ & $\frac{v}{c}^{\rm a}$ & $\alpha$ \\ 
		& & & & $[{\rm meV]}$ & $[{\rm eV}]$ & $[{\rm eV}]$ & $[{\rm eV}]$ & &   \\
		\mr
		${\rm WS_2}$ & $0.26$ & $-0.35$ & $0.39$ & $30$ & $0.42$ & $2.0$ & $0.6$ & $1.7\times 10^{-3}$ & $0.5$\\
		${\rm WSe_2}$ & $0.28$ & $-0.36$ & $0.35$ & $38$ & $0.46$ & $1.7$ & $0.6$ & $1.6\times 10^{-3}$ & $0.6$\\
		\br
		$^{\rm a}$~Ref.~\cite{kdotp}
		\\
		$^{\rm b}$~Ref.~\cite{excitons}
	\end{tabular}
\end{table*}
We note that for the excitons, $\Psi^{\dagger}_{c}(\vec{r}_1)\Psi_v(\vec{r}_2)\rightarrow X^{\dagger}(\vec{R})\phi(\vec{r}_1-\vec{r}_2)$, where $\vec{R}$ is the center-of-mass position of the exciton.
All possible states of the exciton boson operator $X$, can be further classified according to the irreps.
The dark and bright excitonic states transform according to the 
direct product representation of the $c$ and $v$ band states 
$E_1'\otimes E_2'=E\oplus E_3'$. The excitons can further be classified according to the spin projection $S_z$. Due to conservation of spin, the bright state must have $S_z=0$, which corresponds to the excited state transforming according to the $E$ irrep, and only such a combination enters into Eq.~(\ref{eq:rad}). The $(E_3', S_z=0)$ states are the intervalley excitons that are dark due to the momentum mismatch. However, these states can radiatively recombine with the help of phonon emission. According to Table~\ref{prod}, this can be provided by phonons from $E_1'$ and $E_2'$ irreps.
The other exciton doublets, $(E, |S_z|=1)$ and $(E_3',|S_z|=1)$, are absolutely dark due to spin conservation ($\sigma_h$ reflection changes the sign of the spin projection $S_z$.)

According to the sign of the spin-orbit splitting in the c band [see text after Eq.~(1)], $(E_3', S_z=0)$ and $(E, |S_z|=1)$ are the exciton ground states. Therefore, the decay of these states is the bottleneck for the PL. In this case the PL quantum efficiency is determined by the ratio of the radiative and non-radiative Auger rates.

The Auger process is caused by the electron-electron scattering with momentum transfer of the order of inverse lattice constant, therefore in the effective mass approximation it is described by a contact interaction. Table~\ref{c3v''} shows that the only combination allowed by symmetry is 
\begin{eqnarray}
\fl H_{c}=\frac{\alpha \hbar^2}{m_{c'}} \sum_{\sigma,\tau}\int d^2\vec{r}\left( \Psi^{\dagger}_{v\sigma}\Psi^{\dagger}_{c'-\sigma}\Psi_{c-\sigma}\Psi_{c\sigma}\right)_{\vec{r}\tau}+h.c.
\end{eqnarray}
Here, $\alpha$ is a dimensionless parameter computed from the matrix element of the Coulomb interaction in the basis of the density functional theory (DFT) wavefunctions (see Table~\ref{data}).
According to Table~\ref{prod}, the initial dark state exciton $(E_3',S_z=0)$ and the electron $(E_1')$ direct product does not include the final state $c'$ state $E_3'$, making the direct process impossible, hence requiring an extra phonon in the final state. Using Table~\ref{prod}, one finds that this additional phonon in the final state should be the same $\sigma_h$ symmetric ($E_1'$ or $E_2'$) as involved in the radiative process, allowing the direct comparison of the radiative and Auger rates without relying on the knowledge of the strength of the electron-phonon interaction constants\footnote{The process involving the emission of an $E_2'$ phonon mode, which couples to the hole, involves the hole scattering into the lower spin-split $v$ band which results in the appearance of the large $D_{SO}$ spin-orbit splitting in the denominator of the rates. Therefore, we neglect the contribution of the $E_2'$ phonon assisted process to the total rates.}.
%
To consider the electron-phonon interaction only on symmetry grounds, we list in Table~\ref{c3v''} the representations corresponding to the in-plane $D_{xy}$, and out of plane $D_z$ modes in the tripled unit cell, which is needed to describe all $\sigma_h$ symmetric phonon modes in the $\Gamma$ and $K$ points. From the decomposition
$D_{xy}=2E\oplus E_1'\oplus 2E_2'\oplus E_3'$ and
$D_{z}=A_1\oplus E_1'$, we conclude that the existence and number of modes needed to facilitate the processes described are protected by symmetry. To mention, in our earlier studies\cite{danovich} we noticed that the coupling of c band electrons with the homopolar phonon mode $A_1$ is very strong, which hints that the $E_1'$ mode of $D_z$, would be the most relevant for the process.

The Hamiltonian describing the $E_1'$ phonon and its interaction with the $c$ band electrons
\begin{eqnarray}
\fl H_{ph}=\hbar\omega\sum_{\tau}\int d^2{\vec{r}}b_{\tau}^{\dagger}(\vec{r})b_{\tau}(\vec{r})\\\nonumber+g\sum_{\sigma,\tau}\int d^2\vec{r} \left(\Psi^{\dagger}_{c\sigma \tau}\Psi_{c\sigma -\tau}b^{\dagger}_{\tau} +h.c.\right),
\end{eqnarray}
where $b(r)$ is the phonon operator in mode $E_1'$ with energy $\hbar\omega$, and $g$ is the coupling coefficient.
The radiative (with photon line shifted down by $\hbar\omega$ from the dark exciton energy) and non-radiative rates processes are calculated using the Fermi Golden rule, with the quantum mechanical amplitudes shown in Fig.~3. The rates are given by
\begin{eqnarray}
\fl \frac{1}{\tau_r}=\frac{8E_g}{3\hbar}\frac{e^2}{\hbar c} \left(\frac{v}{c}\right)^2\frac{|\phi(0)|^2g^2}{(\Delta_{SO}+\hbar\omega)^2},
\end{eqnarray}
\begin{eqnarray}
\fl \frac{1}{\tau_{A}}=\frac{E_g}{\hbar}\frac{\hbar^2 n_e}{m_{c'} E_g} 
 \frac{ \alpha^2|\phi(0)|^2g^2}{[\Delta_{SO}+\hbar\omega+\frac{m_{c'}}{|m_v|+m_c}\Upsilon]^2},
\end{eqnarray}
where $n_e$ is the electron density.
Taking the ratio of the two rates we obtain
\begin{eqnarray}
 \frac{\tau_r}{\tau_A}=\frac{n_e}{n_e^*},
\end{eqnarray}
where the characteristic density is given by
\begin{eqnarray}
\fl n_e^*=
\frac{8m_{c'}E_g}{3\hbar^2}\left(\frac{v}{\alpha c}\right)^2
\left(\frac{e^2}{\hbar c} \right)\left(1+\frac{\frac{m_{c'}}{|m_v|+m_c}\Upsilon}{\Delta_{SO}+\hbar\omega} \right)^2.
\end{eqnarray}
\begin{figure}[!t]
	\centering
	\includegraphics[width=0.45\textwidth]{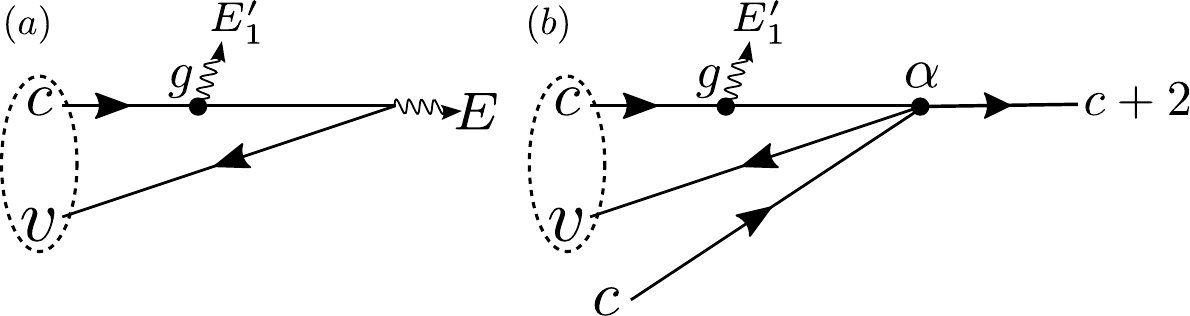}
	\caption{Diagrams for the calculation of the quantum mechanical amplitudes for the (a) phonon assisted radiative process, and (b) phonon assisted Auger process.} 
	\label{fig:fig_amp}
\end{figure}
We emphasize that the latter equation does not involve the unknown electron-phonon coupling constant. Therefore, we can estimate the values of $n_e^*$ for ${\rm WS_2}$ and ${\rm WSe_2}$ based on the parameters of these 2D crystals found in DFT and the experimentally known $E_g$, listed in Table~\ref{data}:
\begin{eqnarray}
\fl n_e^*({\rm WS_2})\sim 10^{10}~{\rm cm^{-2}},
\, n_e^*({\rm WSe_2})\sim 4\times 10^{9}~{\rm cm^{-2}}.
\end{eqnarray}
These electron concentrations which determine the threshold for efficient photoluminescence are remarkably low. This suggests that the proposed mechanism of Auger recombination dominates over the radiative recombination for all realistic structures.

\smallskip
The authors thank T. Heinz, M. Potemski and A. Tartakovski for discussions.
This work was supported by Simons Foundation (IA), ERC Synergy Grant Hetero2D (VF), EC-FET European Graphene Flagship (VZ), EPSRC grant EP/N010345/1 (VF, MD).
\\
\bibliographystyle{iopart-num} 
\bibliography{auger_paper_aleiner_2dmaterials}
	
\end{document}